\documentclass[final]{ws-p8-50x6-00}
\newcommand{\newblock}{}

\begin{document}
\title{Design of a Parallel and Distributed Web Search Engine}

\author{S. Orlando$^{\circ}$, R. Perego$^{*}$, 
F. Silvestri$^*$$^{\bullet}$}
\address{$^{\circ}$Dipartimento di Informatica, Universita Ca' Foscari, 
Venezia, Italy\\
$^{*}$Istituto CNUCE-CNR, Pisa, Italy\\
$^{\bullet}$Dipartimento di Informatica, Universita di Pisa, Italy}

\maketitle

%%%%%%%%%%%%%%%%%%%%%%%%%%%%%%%%%%%%%%%%%%%%%%%%%%%%%%%%%%%%%%%%%%
% Abstract.                                                                    
%%%%%%%%%%%%%%%%%%%%%%%%%%%%%%%%%%%%%%%%%%%%%%%%%%%%%%%%%%%%%%%%%%
\abstracts{ This paper describes the architecture of MOSE ({\em My Own
Search Engine}), a {\em scalable} parallel and distributed engine for
searching the web.  MOSE was specifically designed to efficiently
exploit affordable parallel architectures, such as clusters of
workstations.  Its modular and scalable architecture can be easily
adjusted to fulfill the bandwidth requirements of the application at
hand. Both {\em task-parallel} and {\em data-parallel} approaches are
exploited within MOSE in order to increase the throughput and
efficiently use communication, storing and computational resources.
We used a collection of html documents as a benchmark and conducted
preliminary experiments on a cluster of three SMP Linux PCs.  }
%%%%%%%%%%%%%%%%%%%%%%%%%%%%%%%%%%%%%%%%%%%%%%%%%%%%%%%%%%%%%%%%%%%%%%%%%%%
% Introduction.
%%%%%%%%%%%%%%%%%%%%%%%%%%%%%%%%%%%%%%%%%%%%%%%%%%%%%%%%%%%%%%%%%%%%%%%%%%%
\section{Introduction}
\label{sec:intro}
Due to the explosion in the number of documents available online
today, Web Search Engines (WSEs) have become the main means for
initiating navigation and interaction with the Internet.  Largest WSEs
index today hundreds of millions of multi-lingual web pages containing
millions of distinct terms.  Although bigger is not necessarily
better, people looking the web for unusual (and usual) information
prefer to use the search engines with the largest web coverage. This
forced main commercial WSEs to compete for increasing the indexes.
Since the cost of indexing and searching grows with the size of the
data, efficient algorithms and scalable architectures have to be
exploited in order to manage enormous amount of
information with high throughputs.  Parallel processing thus
become an enabling technology for efficiently searching and retrieving
information from the web.

In this paper we present MOSE, a parallel and distributed WSE able to
achieve high throughput by efficiently exploiting a low cost cluster
of Linux SMPs. Its expansible architecture allows the system to be scaled with
the size of the data collection and the throughput requirements.
Most of our efforts were directed toward increasing {\em query
processing} throughput.  We can think of a WSE as a system with two
inputs and one output. One input is the {\em stream} of queries
submitted by users. The other input is the {\em read-only database},
which contains the index of the document collection. The WSE process
each query of the stream by retrieving from the index the references to
the $l$ most relevant documents. Such set of $l$ references is then
put on the output stream.  
%It is worth considering that the output of each query is, in
%principle, a function of the whole database.
The main parallelization strategies for a WSE are thus:

\begin{enumerate}
\item[\ ] {\em Task parallel}.\ \ Since the various queries can be
processed independently, we can consider {query processing} an
{\em embarrassingly parallel} problem.  We can thus exploit a {\em
processor farm} structure with a mechanism to balance the load
by scheduling the queries among a set of identical workers, each
implementing a sequential WSE.  

\item[\ ] {\em Data parallel}.\ \ The input database is partitioned.
Each query is processed in parallel by several data parallel tasks,
each accessing a distinct partition of the database.  Query processing
is in this case slightly heavier than in the previous case. Each data
parallel task has in fact to retrieve from its own partition the
locally most relevant $l$ references. The final output is obtained by
combining these partial outputs, and by choosing the $l$ references
which globally result to be the most relevant. 

\item[\ ] {\em Task + Data parallel}.\ \ A combination of the above
two strategies.  We have a {\em processor farm}, whose workers are in
turn parallelized using a {\em Data parallel} approach. The farming
structure is used to balance the work among the parallel workers.
\end{enumerate}

The modular architecture of MOSE allowed us to experiment all the
three strategies above.  The third parallelization strategy, which
combines {\em Task} and {\em Data parallelism}, achieved the best
performances due to a better exploitation of memory hierarchies.  

The paper is organized as follow. Section \ref{sec:irs} introduces WSE
and Information Retrieval (IR) principles, and surveys related
work. Section \ref{sec:structure} describes MOSE components, discusses
parallelism exploitation, and shows how MOSE modular and scalable
architecture can be adjusted to fulfill bandwidth requirements.  The
encouraging experimental results obtained on a cluster of three Linux
SMPs are shown in Section \ref{sec:results}, while Section
\ref{sec:conclusions} draws some conclusions.

%%%%%%%%%%%%%%%%%%%%%%%%%%%%%%%%%%%%%%%%%%%%%%%%%%%%%%%%%%%%%%%%%%%%%%%%
% IRSs & web SEs.                                                              
%%%%%%%%%%%%%%%%%%%%%%%%%%%%%%%%%%%%%%%%%%%%%%%%%%%%%%%%%%%%%%%%%%%%%%%%
\section{WSE and IR Principles}
\label{sec:irs}
\begin{figure}
\begin{center}
\epsfxsize=12cm
\epsfbox{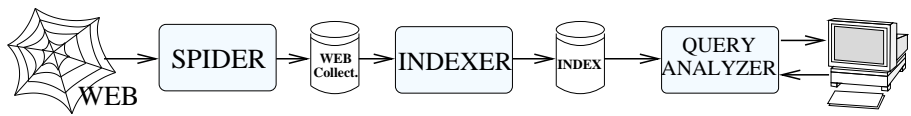}
\end{center}
\caption{Typical organization of a WSE.}
\label{wsestructure}
\end{figure}

A typical WSE (see Figure \ref{wsestructure}) is composed of the {\em
spidering} system, a set of Internet agents which in parallel visit
the web and gather all the documents of interest, and by the IR core
constituted by: (1) the {\em Indexer}, that builds the Index from the
collection of gathered documents, and, (2) the {\em Query Analyzer},
that accepts user queries, searches the index for documents 
matching the query, and return the \emph{references} to these
documents in an understandable form.
Query results are returned to users sorted by \emph{rank}, a kind of
relevance judgment that is an abstract concept largely linked to users
taste.  Ranking is performed on the basis of an IR model that allows
to represent documents and queries, and to measure their similarity.
In general, as the size of the indexed collection grows, a very high
{\em precision} (i.e. number of relevant documents retrieved over the
total number of documents retrieved) has to be preferred even at the
expense of the {\em recall} parameter (i.e. number of relevant
documents retrieved over the total number of relevant documents in the
collection). In other words, since users usually only look at the
first few tens of results, the relevance of these top results is more
important than the total number of relevant documents retrieved.  In
order to grant high precision and computational efficiency, WSEs
usually adopt a simple \emph{Weighted Boolean} IR model enriched with
highly effective ranking algorithms which consider the hyper-textual
structure of web documents\cite{googlearch,mg}.  Moreover, due to its
compactness, most WSEs adopt an \emph{Inverted List} (IL) organization
for the index. An IL stores the relations among a term and the
documents that contain it. The two main components of an IL index are:
(1) the {\em Lexicon}, a lexicographically ordered list of all the
interesting terms contained in the collection, and, (2) the {\em
Postings lists}, lists associated to each term $t$ of the Lexicon
containing the references to all the documents that contain $t$.

Many large-scale WSEs such as {\bf Google}, {\bf Inktomi} and {\bf
Fast}, exploit clusters of low-cost workstation for running their
engines, but, unfortunately, very few papers regard WSE architecture
design\cite{googlearch,fastwp}, since most developments were done
within competitive companies which do not publish technical
details. On the other hand, many researchers investigated parallel
and/or distributed IR
systems\cite{macleod,martinmacleod,burkowski,zhou,lu,cahoon2} focused
on collections of homogeneous documents.
%Proposals of distributed IR systems running on networks of old
%generation servers can be found
%in\cite{macleod,martinmacleod,burkowski}.
%Macleod \emph{et al.}\cite{macleod,martinmacleod} investigated caching
%issues and various index distribution strategies, while
%Burkowski\cite{burkowski} compared two different distributed IR
%systems: in the former functionalities are equally subdivided among a
%server pool, while in the latter servers are specialized.  
Lin and Zhou\cite{zhou} implemented a distributed IR system on a
cluster of workstations, while Lu\cite{lu}, simulated an interesting
distributed IR system on a Terabyte collection, and investigated
various distribution and replication strategies and their impact on
retrieval efficiency and effectiveness.

\begin{figure}
\begin{center}
\epsfxsize=8cm
\epsfbox{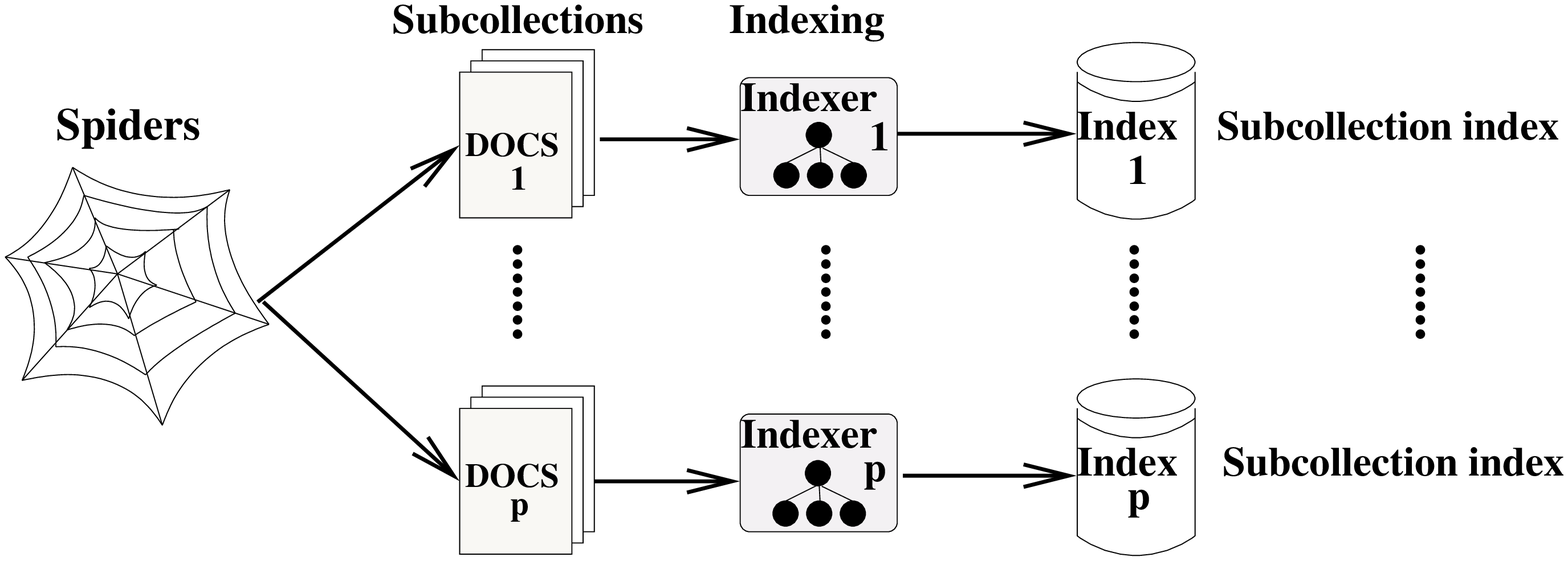}
\end{center}
\caption{Indexing phase.}
\label{documentpartitioning}
\end{figure}

%%%%%%%%%%%%%%%%%%%%%%%%%%%%%%%%%%%%%%%%%%%%%%%%%%%%%%%%%%%%%%%%%%%%%%%%%%
% MOSE Structure.         
%%%%%%%%%%%%%%%%%%%%%%%%%%%%%%%%%%%%%%%%%%%%%%%%%%%%%%%%%%%%%%%%%%%%%%%%%
\section{MOSE Structure}
\label{sec:structure}
\begin{figure}
\begin{center}
\epsfxsize=8cm
\epsfbox{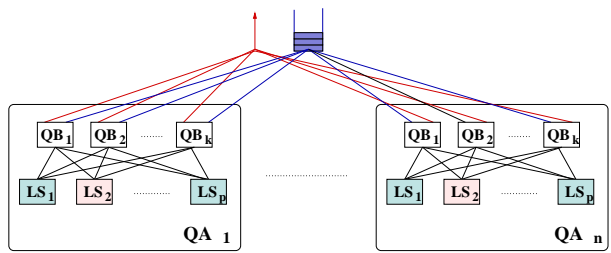}
\end{center}
\caption{Structure of MOSE Query Analyzer.}
\label{distributedMOSE}
\end{figure}

The IR core of MOSE is composed of the {\em Indexer} and the {\em
Query Analyzer} (QA) modules. In this paper we only briefly surveys
indexing issues, and focus our attention on the QA whose
functionalities are carried out by two pools of parallel processes:
{\em Query Brokers} (QBs) and {\em Local Searchers} (LSs).  MOSE
parallel and distributed implementation exploits a data-parallel
technique known as {\em document partitioning} .
%The whole collection of $D$ documents gathered from the web is firstly
%partitioned into $p$ subcollections of $D/p$ documents.
The spidering phase returns $p$ subcollections of documents with
similar sizes.  The subcollections are then indexed independently and
concurrently by $p$ parallel {\em Indexers} (see Figure
\ref{documentpartitioning}). The result of the indexing phase is a set
of $p$ different indexes containing references to disjoint sets of
documents. The $p$ indexes are then taken in charge by a data-parallel
QA whose task is to resolve user queries on the whole collection. To
this end the QA uses $k$ QBs and $p$ LSs.  The $k$ QBs run on a front-end
workstation, and fetch user queries from a shared message queue. Every
fetched query is then broadcast to the associated $p$ LSs ({\em
workers}), possibly running on different workstations.  The $p$ LSs
satisfy the query on the distinct subindexes, and return to the QB
that submitted the query the first $l$ references to most relevant
documents contained within each subcollection. The QB waits for all
the $l \cdot p$ results and chooses among them the $l$ documents with
the highest ranks.  Finally, such results are returned to the
requesting user. Figure \ref{distributedMOSE} shows the logic
structure of the MOSE architecture. A QB, along with the $p$
associated LSs, implements a {\em data parallel} worker which
concurrently serve the user queries.  In order to manage concurrently
more queries and to better exploit LSs' bandwidth, $k$ QBs are
introduced within a QA. System performances can be furthermore
increased by replicating the QA in $n$ copies.
%In this case, two LS$_i$ belonging to distinct QAs may me mapped to
%the same workstation (if the computational bandwidth of the
%workstation is under utilized) or to distinct workstations (to improve
%the availability and the throughput of the whole system).  

All the parallelization strategies depicted in Section \ref{sec:intro}
can be thus realized by choosing appropriate values for $n$, $k$, and
$p$. A pure task parallel approach corresponds to $p=1$, while $n>1$
and/or $k>1$.  By choosing $p>1$, $n=1$ and $k=1$ we obtain a pure
data-parallel implementation. A hybrid task + data parallel strategy
is finally obtained for $p>1$, while $n>1$ and/or $k>1$.

\medskip
\noindent
{\bf Indexer.}  The Indexer has the purpose of building the index from
the gathered web documents. The indexing algorithm used is a parallel
version of the \emph{Sort Based} algorithm which is very efficient on
large collections due to the good compromise between memory and I/O
usage\cite{mg}. Moreover, the index built is \emph{Full Text} and
\emph{Word Based}. The Lexicon is compressed exploiting the common
prefixes of lexicographically ordered terms ({\em Shared Prefix
Coding}), while the Postings lists are compressed by using the
\emph{Local Bernoulli} technique\cite{mg}. MOSE parallel Indexer
exploits the \emph{master/worker} paradigm and standard Unix SysV
communication mechanisms (i.e. message queues). Since each
subcollection of web documents is indexed independently (and
concurrently on different workstations), the current Indexer
implementation exploits parallelism only within the same SMP
architecture.  The master process scans the subcollection, and sends
the reference to each document (i.e. the file offset) along with a
unique document identifier to one of the worker processes on a
self-scheduling basis. The workers independently read each assigned
document from the disk and indexes it. When all documents have been
processed, the workers write their local indexes to the disk, and
signal their completion to the master. At this point the master merges
the local subindexes in order to create a single index for the whole
subcollection.  A distributed implementation of the Indexer could be
easily derived, but should require all the processing nodes to
efficiently access the disk-resident subcollection, and that at least
a single node can access all the subindexes during the merging phase.

%%%%%%%%%%%%%%%%%%%%%%%%%%%%%%%%%%%%%%%%%%%%%%%%%%%%%%%%%%%%%%%%%%%%%%%%
% Query Broker. 
%%%%%%%%%%%%%%%%%%%%%%%%%%%%%%%%%%%%%%%%%%%%%%%%%%%%%%%%%%%%%%%%%%%%%%%%
\medskip
\noindent
{\bf Query Broker.}
Each QB loops performing the following actions:

\noindent
{\em Receipt and broadcasting of queries.}
Independently from the mechanism exploited to accept user
queries (e.g., CGI, fast CGI, PHP, ASP), user queries are inserted
in a SysV message queue shared among all the QBs.  Load balancing
is accomplished by means of a \emph{self scheduling}
policy: free QBs access the shared queue and get the first available
query.  Once a query is fetched, the QB broadcasts it
to its $p$ LSs by means of an MPI asynchronous communication.

\medskip
\noindent
{\em Receipt and merge of results.}  The QB then nondeterministically
receives the results from all the LSs (i.e., $p$ lists ordered by
rank, of $l$ pairs {\em document identifier}, and associated {\em rank
value}) . The final list of the $l$ results with the highest ranks is
than obtained with a simple $O(l)$ merging algorithm.

\medskip
\noindent
{\em Answers returning.}  The list of $l$ results is finally returned to
the CGI script originating the query that transforms document
identifiers into {\tt URLs} with a short abstract associated, and
builds the dynamic {\tt html} page returned to the requesting user.

%%%%%%%%%%%%%%%%%%%%%%%%%%%%%%%%%%%%%%%%%%%%%%%%%%%%%%%%%%%%%%%%%%%%
% Local Searcher.                                                              
%%%%%%%%%%%%%%%%%%%%%%%%%%%%%%%%%%%%%%%%%%%%%%%%%%%%%%%%%%%%%%%%%%%%
\medskip
\noindent
{\bf Local Searcher.}  LSs implement the IR engine of MOSE. Once a
query is received, the LS parses it, and searches the Lexicon for each
terms of the query. Performance of term searching is very important
for the whole system and are fully optimized. An efficient binary
search algorithm is used at this purpose, and a \emph{Shared Prefix
Coding} technique is used to code the variable length terms of the
lexicographically ordered Lexicon without wasting space\cite{mg}.
Minimizing the size of the Lexicon is very important: a small Lexicon
can be maintained in core with obvious repercussions on searching
times.  LS exploit the Unix {\em mmap} function to map the Lexicon
into memory. The same function also allows an LS to share the Lexicon
with all the other LS that run on the same workstation and process the
same subcollection.  Once a term of the query is found in the Lexicon,
the associated posting list is retrieved from the disk,
decompressed, and written onto a stack.  The LS then processes {\em
bottom-up} query boolean operators whenever their operands are
available onto the top of the stack.  When all boolean operators have
been processed, the top of the stack stores the final list of
results. The $l$ results with the highest ranks are then selected in
linear time by exploiting a \emph{max-heap} data
structure\cite{mg}. Finally, the $l$ results are communicated to the
QB that submitted the query.

%%%%%%%%%%%%%%%%%%%%%%%%%%%%%%%%%%%%%%%%%%%%%%%%%%%%%%%%%%%%%%%%%%%%%%
% Experimental Results.
%%%%%%%%%%%%%%%%%%%%%%%%%%%%%%%%%%%%%%%%%%%%%%%%%%%%%%%%%%%%%%%%%%%%%%
\section{Experimental Results}
\label{sec:results}
We conducted our experiments on a cluster of three SMP Linux PCs
interconnected by a switched Fast Ethernet network. Each PC is
equipped with two 233MHz PentiumII processors, 128 MBytes of RAM, and
an ULTRA SCSI II disk.  We indexed 750.000 multi-lingual html documents
contained in the CDs of the web track of the TREC Conference and we
built both a monolithic index ($p=1$) and a partitioned one ($p=2$).
The monolithic index contains 6.700.000 distinct terms and has a size
of 0.96 GBytes (1.7 GBytes without compression), while each one of the
two partitions of the partitioned index occupy about 0.55 GBytes.  The
queries used for testing come from an actual query log file provided
by the Italian WEB Search Company IDEARE S.p.A.

We experimented Task-Parallel (TP), and hybrid (TP + DP)
configurations of MOSE.  We mapped all the QBs on a single
workstation, while the LSs were placed on one or both the other
machines.  Independently of the configuration used (one or two index
partitions), two QBs were introduced ($k=2$).
\begin{figure}
\centerline{
\includegraphics[angle=-90,width=6cm]{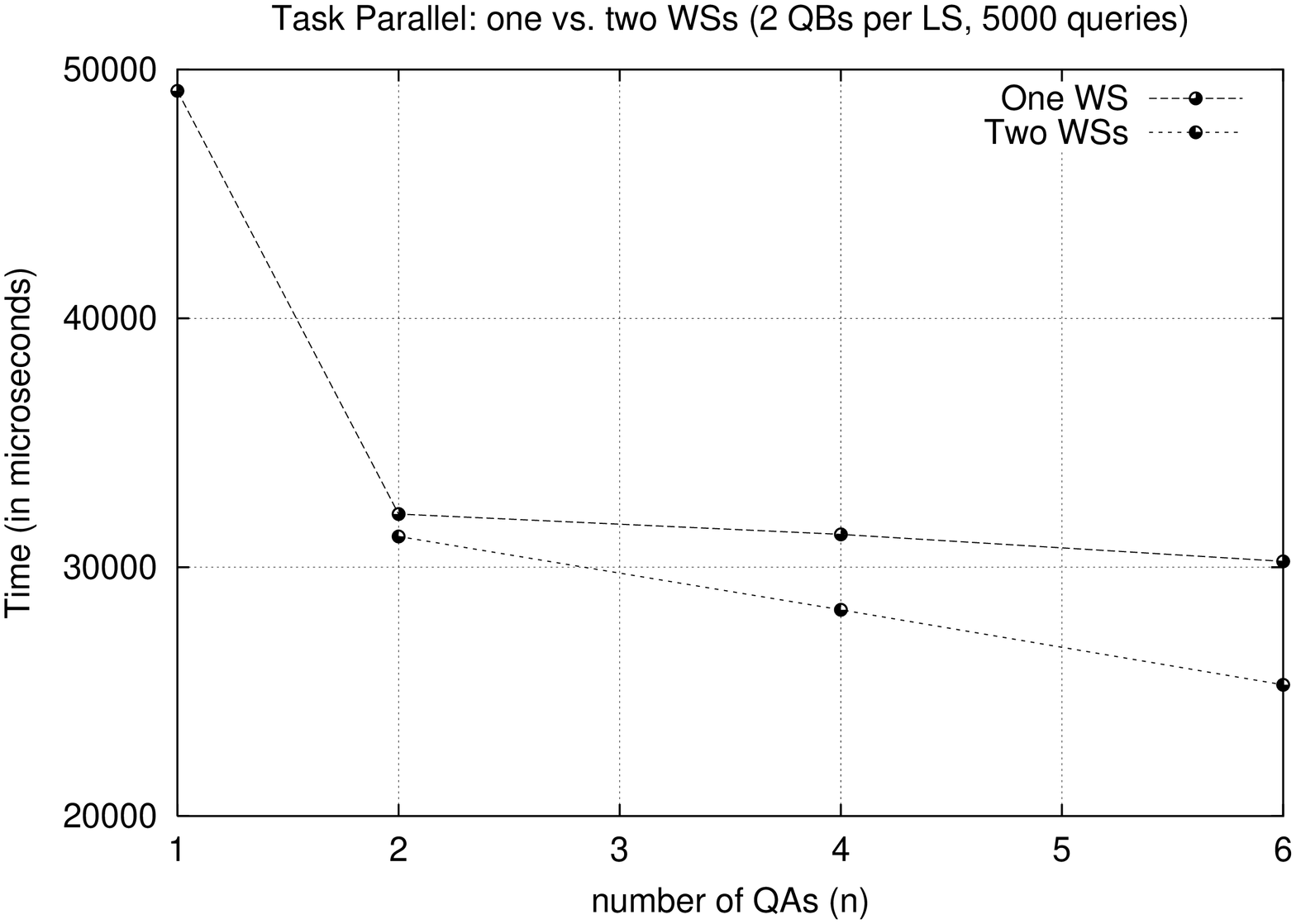}
\includegraphics[angle=-90,width=6cm]{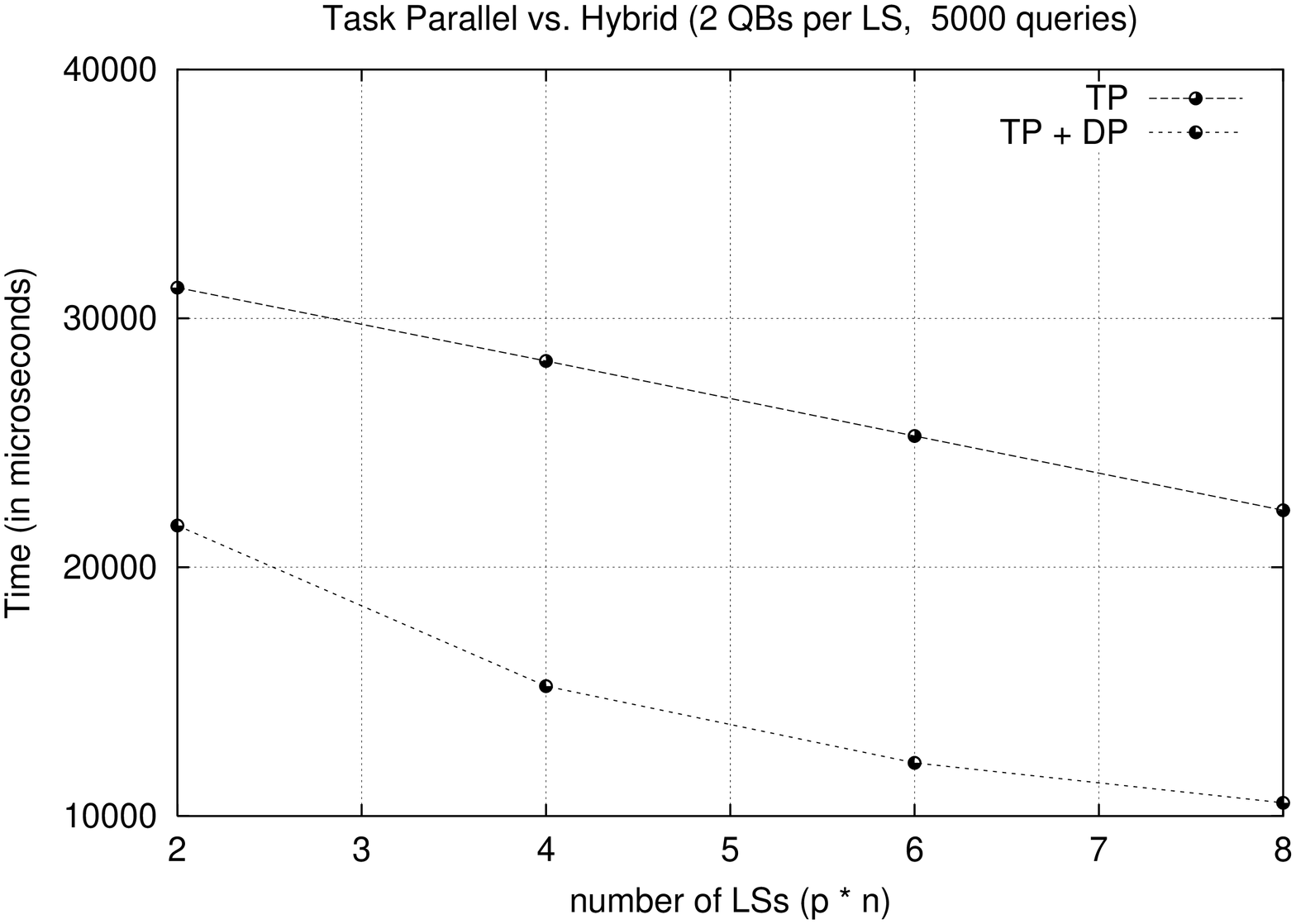}
}
\centerline{(a) \hspace{5.5cm} (b)}
\caption{Results of the experiments conducted.}
\label{onetwo}
\end{figure}
Figure~\ref{onetwo}.(a) reports the average elapsed times, i.e. the
inverse of the throughput, required to process each one of 5000
queries for the TP case ($p = 1$) as a function of $n$, i.e. the
number of QAs exploited. The two curves plotted refer to the cases
where the LSs were mapped on one or two SMP machines.  We can see that
when two QAs are used they can be almost indifferently placed on one
or two SMP machines, thus showing the efficacy of the sharing
mechanisms used. On the other hand, as we increase the number of QAs,
the difference between exploiting one or two machines increases as
well. We can also observe that it is useful to employ more QAs than 
the available processors.

Figure~\ref{onetwo}.(b) compares the TP solution with the hybrid one
(TP + DP). Testing conditions were the same as the experiment
above. In the case of the hybrid configuration, all the LSs associated
with the same partition of the index were placed on the same
workstation in order to allow the LSs to share the lexicon data
structure. The better performance of the hybrid approach is
evident. Superlinear speedups were obtained in all the TP + DP
tests. They derive from a good exploitation of memory hierarchies,
in particular of the buffer cache which virtualize the accesses to the
disk-resident posting lists.
%%%%%%%%%%%%%%%%%%%%%%%%%%%%%%%%%%%%%%%%%%%%%%%%%%%%%%%%%%%%%%%%%%%%%
% Conclusions.
%%%%%%%%%%%%%%%%%%%%%%%%%%%%%%%%%%%%%%%%%%%%%%%%%%%%%%%%%%%%%%%%%%%%%
\section{Conclusions}
\label{sec:conclusions}
We have presented the parallel and distributed architecture of MOSE,
and discussed how it was designed in order to efficiently exploit
low-cost clusters of workstations.  We reported the results of
preliminary experiments conducted on three SMP workstations. The
results highlighted the greater performances resulting from exploiting
a hybrid {\em Task + Data} parallelization strategy over a pure {\em
Task-parallel} one. There are a lot of important issues we plan to
investigate in the near future.  The most important is performing an
accurate testing of MOSE on larger clusters and document collections
in order to analyze in greater detail the scalability of the different
parallelization strategies. Fastest interconnection network such as
Myrinet have also to be tested. Moreover, we are interested to study
query locality and the effectiveness of caching their results within
QBs, and ``supervised'' document partitioning strategies aimed at
reducing the number of index partitions needed to satisfy each query.

%\bibliography{ir}
\end{document}